\documentclass[%
 reprint,
showpacs,%preprintnumbers,
 amsmath,amssymb,
 aps,
prb,
]{revtex4-1}

\usepackage{graphicx}% Include figure files
\usepackage{dcolumn}% Align table columns on decimal point
\usepackage{bm}% bold math
\usepackage{multirow}
\usepackage{tikz}
\usepackage{pifont}
\begin{document}

\preprint{APS/123-QED}

\title{Electronic structure and phonon instabilities in the vicinity of the structural transition and superconductivity of (Sr,Ca)$_3$Ir$_4$Sn$_{13}$}% Force line breaks with \\
%\thanks{A footnote to the article title}%

\author{D.A. Tompsett}
% \altaffiliation[Present Address]{Physics Department, XYZ University.}%Lines break automatically or can be forced with \\
\email{dt331@bath.ac.uk}
\affiliation{Department of Chemistry, University of Bath, Bath BA2 7AY, UK.}
%\\ This line break forced with \textbackslash\textbackslash
%
%\affiliation{Department of Chemistry, University of Cambridge, Cambridge CB2 1EW, UK.}

\date{\today}% It is always \today, today,
             %  but any date may be explicitly specified

\begin{abstract}
The nature of the lattice instability connected to the structural transition and superconductivity of (Sr,Ca)$_3$Ir$_4$Sn$_{13}$ is not yet fully understood. In this work density functional theory (DFT) calculations of the phonon instabilities as a function of chemical and hydrostatic pressure show that the primary lattice instabilities in Sr$_3$Ir$_4$Sn$_{13}$ lie at phonon modes of wavevectors $\mathbf{q}=(0.5,0,0)$ and $\mathbf{q}=(0.5,0.5,0)$. Following these modes by calculating the energy of supercells incorporating the mode distortion results in an energy advantage of -14.1 meV and -9.0 meV per formula unit respectively. However, the application of chemical pressure to form Ca$_3$Ir$_4$Sn$_{13}$ reduces the energetic advantage of these instabilities, which is completely removed by the application of a hydrostatic pressure of 35 kbar to Ca$_3$Ir$_4$Sn$_{13}$. The evolution of these lattice instabilities is consistent with experimental phase diagram. The structural distortion associated with the mode at $\mathbf{q}=(0.5,0.5,0)$ produces a distorted cell with the same space group symmetry as the experimentally refined low temperature structure. Furthermore, calculation of the deformation potential due to these modes quantitatively demonstrates a strong electron-phonon coupling. Therefore, these modes are likely to be implicated in the structural transition and superconductivity of this system. 
%\begin{description}
%\item[Usage]
%Secondary publications and information retrieval purposes.
%\item[PACS numbers]
%May be entered using the \verb+\pacs{#1}+ command.
%\item[Structure]
%You may use the \texttt{description} environment to structure your abstract;
%use the optional argument of the \verb+\item+ command to give the category of each item. 
%\end{description}
\end{abstract}

% PACS, the Physics and Astronomy
                             % Classification Scheme.
%Use showkeys class option if keyword
                              %display desired
\maketitle

%\tableofcontents

\section{\label{sec:introduction}Introduction}
% Jazz this up. Intro in Kudo is good. Its Ref 20, subedi and Singh, may be good too. BaNi2As2 and CaC6 both give good cases of enhanced superconductivity at structural phase transition. Ours is different since still cubic structure. Swee suggests to big up superconductivity. Structural instabilities have also recently been found in BaNi2As2 and bismuth oxides. Superconductivity has recently been shown to not be accompanied by spin fluctuations in muSR measurements and also to be fully gapped. Therefore there is a pressing need to understand the origin of the T*!! He also mentioned something about a first synthesis by espinoza. See also the recent paper by the Taiwanese group on La3Co4Sn13.
Understanding the mechanisms of superconductivity, despite the documentation of this phenomenon for over a century, remains a pressing challenge to condensed matter physics. Progress, particularly in enhancing the superconducting transition temperature ($T_c$) may drive future technological innovation in energy storage, energy transmission and information devices. Enhanced $T_c$ is often associated with structural phase transitions, as observed in copper oxides\cite{Lee1, Reznik1}, iron pnictides\cite{Yoshizawa1, Niedziela1} and CaC$_6$\cite{Gauzzi1}, and recently the action of soft phonons has been implicated in the superconductivity of BaNi$_2$As$_2$\cite{Kudo1} ($T_c=3.3$ K) and the RO$_x$F$_{1-x}$BiS$_2$ ($T_c=10$ K) compounds\cite{Yildirim2}.
The discovery of superconductivity in (Sr,Ca)$_3$Ir$_4$Sn$_{13}$, with $T_c$ up to 8.9 K, presents a new opportunity to enhance our understanding of the interplay between superconductivity and structural instabilities in a system without an associated magnetic effect\cite{Espinosa1, Yang1}. Furthermore, a comprehensive study of the phase diagram of (Sr,Ca)$_3$Ir$_4$Sn$_{13}$ has demonstrated the ability to tune its observed structural distortion to a zero temperature quantum phase transition\cite{Klintberg1}. Consequently, there is a driving need to understand the underlying nature of the structural transition and soft phonons in (Sr,Ca)$_3$Ir$_4$Sn$_{13}$, which we address here by a detailed first-principles study.

%Phase transitions that are tuned to zero temperature, such as the structural distortion in (Sr,Ca)$_3$Ir$_4$Sn$_{13}$, have proved fertile ground for novel physics including superconductivity\cite{Mathur1}, non-Fermi liquid metallicity\cite{Smith1} and superfluid phenomena. A deeper understanding of the physics active in these systems is of potential importance to materials science\cite{Coleman1}. This importance is due to the tendency for materials on the border of these phase transitions to reorder into new stable phases of matter.

\begin{figure}
\includegraphics[width=0.525\textwidth,angle=0]{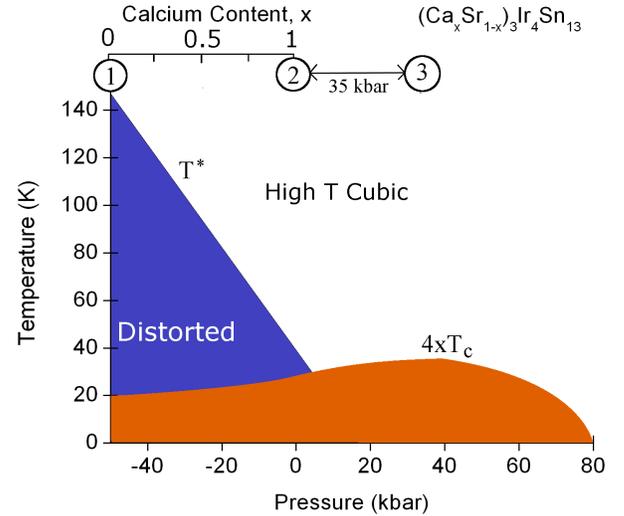}% Here is how to import EPS art
\caption{\label{fig:phaseDiagram} (Color online) Experimental phase diagram for (Ca$_{x}$Sr$_{1-x}$)$_3$Ir$_4$Sn$_{13}$ from Ref.\cite{Klintberg1}. Constructed by placeing $x=0$, 0.5 and 0.75 at -52, -26 and -13 kbar (c.f. top axis). The circled numbers correspond to the position of the three calculations in Table~\ref{table:energiesEvolution} in the composition-pressure space.}
\end{figure}

Recent experimental investigations\cite{Klintberg1, Yang1} into Sr$_3$Ir$_4$Sn$_{13}$ have demonstrated the presence of superconductivity below $T_{c}=5$ K as well as distinct anomalies in the electrical resistivity and magnetic susceptibility at $T^{*} \sim$ 147 K. X-ray diffraction measurements indicate that the anomaly is associated with a structural transition to a superlattice structure. The transition has been traced with both chemical and hydrostatic pressure to elucidate the phase diagram in Fig.~\ref{fig:phaseDiagram}. The application of chemical pressure by alloying calcium in (Ca$_{x}$Sr$_{1-x}$)$_3$Ir$_4$Sn$_{13}$ results in the suppression of the anomaly to $T^{*}=33$ K in Ca$_3$Ir$_4$Sn$_{13}$ and enhancement of the superconducting transition to $T_{c}=7$ K. Proceeding beyond the chemical pressure regime by the application of hydrostatic pressure upon Ca$_3$Ir$_4$Sn$_{13}$ results in suppression of the structural transition towards a zero temperature quantum phase transition at approximately 18 kbar. A superconducting dome extends beyond this pressure and peaks at 40 kbar with a transition temperature of $T_{c}=8.9$ K.

\begin{figure}
\includegraphics[width=0.375\textwidth,angle=0]{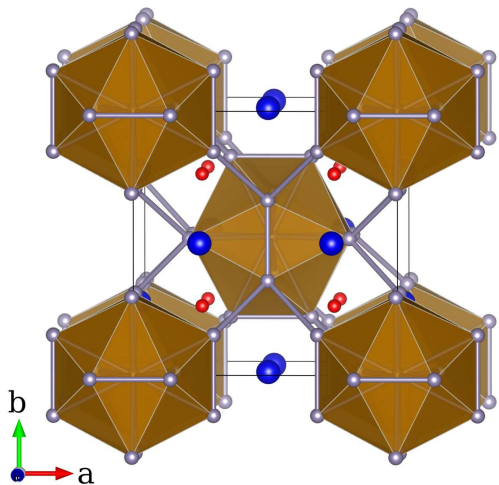}% Here is how to import EPS art
\caption{\label{fig:unitCell} (Color online) Crystal structure of the $I$-phase of (Sr,Ca)$_3$Ir$_4$Sn$_{13}$. The large icosahedra are formed by Sn$_{12}$ units with a further Sn atom at their center. The large blue spheres indicate the (Sr,Ca) position and the small red spheres are the Ir.}
\end{figure}

(Sr,Ca)$_3$Ir$_4$Sn$_{13}$ is only a small part of the $R_3T_4X_{13}$ general stoichiometry, where $R$ is an alkaline-earth or rare-earth element, $T$ is a transition metal and $X$ is a group-IV element. Many members of this family of materials have been observed to undergo a structural distortion to a low temperature structure, the $I'$ phase, which reduces the symmetry of the simple cubic parent structure ($I$ phase, $Pm\bar{3}m$). A series of diffraction studies have not produced agreement in describing the low temperature $I'$ phase\cite{Nagoshi1, Hodeau1, Miraglia1, Bordet1}. The structure of (Sr,Ca)$_3$Ir$_4$Sn$_{13}$ in the high temperature $I$-phase is shown in Fig.~\ref{fig:unitCell} and is dominated by Sn$_{12}$ icosahedra that are joined by triangular prismal IrSn$_6$ units. It is the transition from the $I$ phase at elevated temperatures to the $I'$ that is thought to be the origin of the anomalies at $T^{*}$ in (Sr,Ca)$_3$Ir$_4$Sn$_{13}$. Given the presence of superconductivity and its intriguing interplay with the structural transition it is timely to apply DFT to study the evolution of the phase diagram in (Sr,Ca)$_3$Ir$_4$Sn$_{13}$. In this work we show that the evolution of the phase diagram in Fig.~\ref{fig:phaseDiagram} may be explained by the pressure dependence of the phonon instabilities from DFT.

%\begin{figure}
%\includegraphics[scale=0.2,angle=0]{POSCAR_fixedRadiiColour_labelled}% Here is how to import EPS art
%\caption{\label{fig:unitCell} (Color online) Crystal structure of the $I$-phase of (Sr,Ca)$_3$Ir$_4$Sn$_{13}$. The large icosahedra are formed by Sn$_{12}$ units with a further Sn atom at their center. The large blue spheres indicate the (Sr,Ca) position and the small red spheres are the Ir. The icosahedra are linked by triangular prisms that surround the Ir.}
%\end{figure}

% The Fermi surface is dominated by states of mixed Ir-\textit{d} and Sn-\textit{p} character, but shows comparatively little pressure dependence compared to the phonon spectrum. These results suggest that while coupling to the electronic structure must be important to the superconductivity, the evolution of the phonon spectrum dominates the form of the temperature-pressure phase diagram.

 %This family of materials contains the superconducting and magnetic stannides\cite{Remeika1, Espinosa1}, Ce-based Kondo lattice systems\cite{Sato2}, and thermoelectrics\cite{Strydom1}. Yet, there is a large structural composition space in the neighborhood of (Sr,Ca)$_3$Ir$_4$Sn$_{13}$ where its important physics may be explored. 

\section{\label{sec:results}Results and Discussion}
The electronic structure has been calculated using the Local Density Approximation (LDA). The VASP\cite{Kresse1} code was employed for the calculation of the phonon spectrum in conjuntion with Phonopy\cite{phonopy1}. PAW potentials were employed with a cutoff for the planewave basis set of 350 eV. A minimum of 4$\times$4$\times$4 $k$-point grid was used to achieve good convergence of the force constants. The all-electron full-potenial code Wien2k\cite{Blaha1} was also employed to verify key results. Here RK$_{\textnormal{max}}$ was set to 7.0 and the radii of the muffin tins was 2.5 $a_0$ for (Sr,Ca), 2.5 $a_0$ for Ir and 2.25 $a_0$ for Sn. A minimum of 4000 k-points was used in the full Brillouin zone to obtain an accurate density of states and potential for the calculcation of the bandstructure.

\subsection{\label{subsec:electronic}Electronic Structure}
\begin{figure}
\includegraphics[scale=0.42,angle=0]{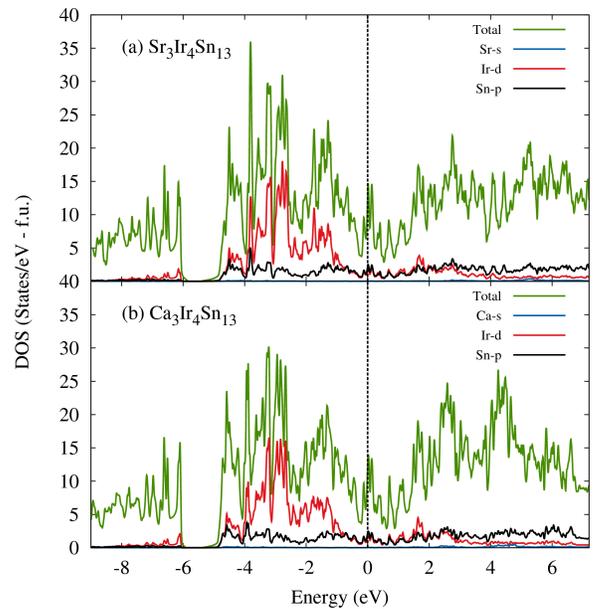}% Here is how to import EPS art
\caption{\label{fig:DOS} (Color online) Density of states of (a) Sr$_3$Ir$_4$Sn$_{13}$ and (b) Ca$_3$Ir$_4$Sn$_{13}$ (f.u.$\equiv$formula unit.).}
\end{figure}

The intricate phase diagram of (Sr,Ca)$_3$Ir$_4$Sn$_{13}$ is evidence of a delicate interplay between the electronic and lattice degrees of freedom in this system. We have calculated the electronic structure for Sr$_3$Ir$_4$Sn$_{13}$ and Ca$_3$Ir$_4$Sn$_{13}$ in the simple cubic $I$-phase using the experimental lattice parameters\cite{Klintberg1, Yang1} of $9.7968(3)$ \AA~and 9.7437 \AA~respectively. The smaller lattice parameter of Ca$_3$Ir$_4$Sn$_{13}$ corresponds to an effective chemical pressure and, as shown in the phase diagram of Fig.~\ref{fig:phaseDiagram}, this material lies near to the critical pressure at which  $T^{*}$ extrapolates to zero temperature. The density of states (DOS) for both materials is shown in Fig.~\ref{fig:DOS}.

Both materials possess a large density of states, approximately 12.5 states/eV per formula unit, at the Fermi level which is indicative of good metallic behavior. The overall appearance of the DOS for the two systems is similar. Projections of the DOS inside the muffin tin spheres for the $s$, $p$ and $d$ angular momenta indicate that the states near the Fermi level are primarily Ir-$d$ and Sn-$p$ contributions. This reflects the crystal structure shown in Fig.~\ref{fig:unitCell} where IrSn$_6$ triangular prisms connect the Sn$_{12}$ icosahedra. The majority of the Ir-$d$ states are loosely bound in a broad structure between -5  and -1 eV below the Fermi level.

Attempts were made to stabilize spin polarized states in both Sr$_3$Ir$_4$Sn$_{13}$ and Ca$_3$Ir$_4$Sn$_{13}$ in ferromagnetic and various antiferromagnetic spin configurations. However, all calculations with the LDA exchange correlation functional, as well as tests with Generalized Gradient Approximations\cite{Perdew1}, resulted in non-magnetic solutions. The lack of magnetism in this system is consistent with the low experimental magnetic susceptibilties and the absence of anisotropy in the susceptibility near $T^{*}$\cite{Klintberg1}, as well as with recent $\mu$SR measurements\cite{Gerber1}. Consequently, in the remainder of this work all reported calculations are non-magnetic.

We will return to discuss further features of the electronic structure, but given the importance of the structural transition at $T^{*}$ we first turn our attention to the phonon driven instabilities.

\subsection{Structural Instabilities}
\begin{figure}
\includegraphics[scale=0.65,angle=0]{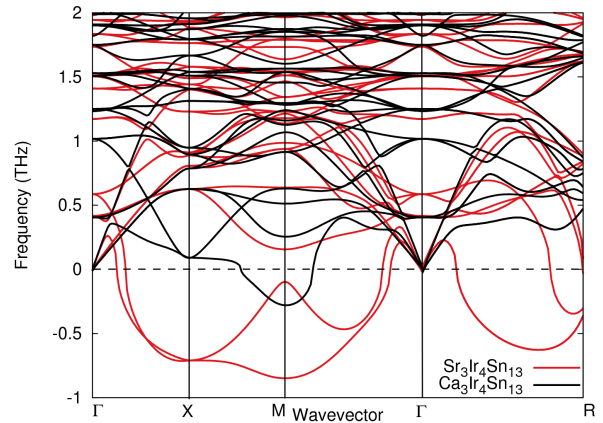}% Here is how to import EPS art
\caption{\label{fig:combinedPhononDispersion} (Color online) Calculated phonon dispersion for Sr$_3$Ir$_4$Sn$_{13}$ and Ca$_3$Ir$_4$Sn$_{13}$ in the $I$-phase. Only low energy frequencies are shown since we are most interested in the evolution of soft modes.}
\end{figure}

The presence of a structural transition from the $I$-phase has been observed in several members of the $R_3T_4X_{13}$ family. However investigations have met with difficulty in indexing the low temperature $I'$ structure\cite{Hodeau1, Miraglia1}. The mapping of the temperature-pressure phase diagram of (Sr,Ca)$_3$Ir$_4$Sn$_{13}$ indicates a composition space in which the structural instability may be traced. The phonon dispersion for the $I$-phase of both Sr$_3$Ir$_4$Sn$_{13}$ and Ca$_3$Ir$_4$Sn$_{13}$ has been calculated and the results are shown in Fig.~\ref{fig:combinedPhononDispersion}. There is a clear distinction between the phonon dispersion in the two materials. Sr$_3$Ir$_4$Sn$_{13}$ possesses imaginary phonon frequencies that indicate the presence of lattice instabilities towards a displacive phase transition. Strong instabilities lie at the \textbf{X} point of the Brillouin zone corresponding to $\mathbf{q}=(0.5,0,0)$ and \textbf{M} point corresponding to $\mathbf{q}=(0.5,0.5,0)$. A further, but weaker, instability is also present at the \textbf{R} point ($\mathbf{q}=(0.5,0.5,0.5)$). 

In the case of Ca$_3$Ir$_4$Sn$_{13}$, which lies much nearer the zero temperature end point for the structural transition, only a single imaginary mode remains at the \textbf{M} point. The lowest mode at \textbf{X} is now real, but with a soft, low frequency. The lowest mode at \textbf{R} has also clearly become a real oscillatory frequency in Ca$_3$Ir$_4$Sn$_{13}$ with significantly higher energy. The clear evolution of these modes as we move from considering Sr$_3$Ir$_4$Sn$_{13}$ to Ca$_3$Ir$_4$Sn$_{13}$ is in concert with the suppression of $T^{*}$ in the phase diagram of Fig.~\ref{fig:phaseDiagram}. In particular the near zero frequency of both the real mode at \textbf{X} and imaginary frequency at \textbf{M} in Ca$_3$Ir$_4$Sn$_{13}$ is consistent with the proximity of this material to the zero temperature end point for the structural transition. At the end point it is expected that the energy scale for the structural transition becomes very small and the associated phonon modes will soften.

It is important to note that the phonon spectrum is calculated in the harmonic approximation. Consequently anharmonic effects may impact the stability of structural distortions associated with these modes. To quantify the energetic advantage of the phonon instabilities at \textbf{X}, \textbf{M} and \textbf{R} the distortion due to these modes has been incorporated into supercells.

In Table~\ref{table:energiesEvolution} the energy of structurally minimized supercells incorporating distortions due to the lowest energy phonon modes at \textbf{X}, \textbf{M} and \textbf{R} are compared to the energy of the undistorted $I$-phase structure. For Sr$_3$Ir$_4$Sn$_{13}$, denoted \ding{192} (c.f. Fig~\ref{fig:phaseDiagram}), the distortions due to the imaginary phonon modes at \textbf{X} and \textbf{M} deliver energy advantages of 14.1 meV and 9.0 meV per formula unit respectively. These energies correspond to approximately 163 K and 105 K which are in reasonable correspondence with the experimental $T^{*}=147$ K for this material. Sr$_3$Ir$_4$Sn$_{13}$ is also unstable towards the phonon mode at \textbf{R}, but with significantly smaller energy advantage of just 4.0 meV per formula unit.

Table~\ref{table:energiesEvolution} also shows the energy of lattice distortions due to the phonon modes in Ca$_3$Ir$_4$Sn$_{13}$, denoted \ding{193} (c.f. Fig~\ref{fig:phaseDiagram}). The energetic advantage due to distortions at \textbf{X} and \textbf{M} are now just 4.6 meV and 2.1 meV per formula unit, which is consistent with the lower $T^{*}\approx 33$ K in Ca$_3$Ir$_4$Sn$_{13}$. The distortion at \textbf{R} is no longer stable and relaxes back to become the undistorted $I$-phase. 

\begin{table*}
\caption{\label{table:energiesEvolution} Calculated stabilization energies of structural distortions due to soft modes relative to the high temperature $I$-phase. The distorted structures for each wavevector are obtained by following the phonon modes of low frequencies in the phonon dispersion of Fig.~\ref{fig:combinedPhononDispersion}. Circle designations \ding{192}, \ding{193} and \ding{194} correspond to Fig~\ref{fig:phaseDiagram}.}

  \begin{tabular*}{0.7\textwidth}{@{\extracolsep{\fill}}cccc}
    \hline
     Distortion  & \ding{192} Sr$_3$Ir$_4$Sn$_{13}$ & \ding{193} Ca$_3$Ir$_4$Sn$_{13}$ & \ding{194} Ca$_3$Ir$_4$Sn$_{13}$ $P=35$ kbar \\
    \hline
    Undistorted $I$-phase    & 0      & 0    & 0 \\
    \textbf{X} $\mathbf{q}=(0.5,0,0)$   & -14.1  & -4.6 & 0 \\
    \textbf{M} $\mathbf{q}=(0.5,0.5,0)$ & -9.0   & -2.1 & 0 \\
    \textbf{R} $\mathbf{q}=(0.5,0.5,0.5)$ & -4.0 & 0    & 0 \\
    \hline

  \end{tabular*}

\end{table*}

In order to trace the evolution of the structural instabilities across the extrapolated zero temperature end point for $T^{*}$ the energies have also been determined for a cell of Ca$_3$Ir$_4$Sn$_{13}$ subject to a pressure of 35 kbar\cite{pressureMethod}, denoted \ding{194} (c.f. Fig~\ref{fig:phaseDiagram}). At this pressure, 17 kbar beyond the extrapolated zero temperature end point, none of the mode distortions give an energetic advantage as shown in the final column of Table~\ref{table:energiesEvolution}. This gives strong evidence that it is indeed these modes that are responsible for the structural distortion associated with $T^{*}$. The presence of the superconducting dome in this region makes the phonon modes, particularly at \textbf{X} and \textbf{M}, strong candidates to be considered in the mechanism for the superconductivity in this system. The presence of these very soft modes also suggests that the structural phase transition may be second order, however a definitive confirmation of this will require high quality specific heat measurements.

\begin{figure}
\includegraphics[width=0.475\textwidth,angle=0]{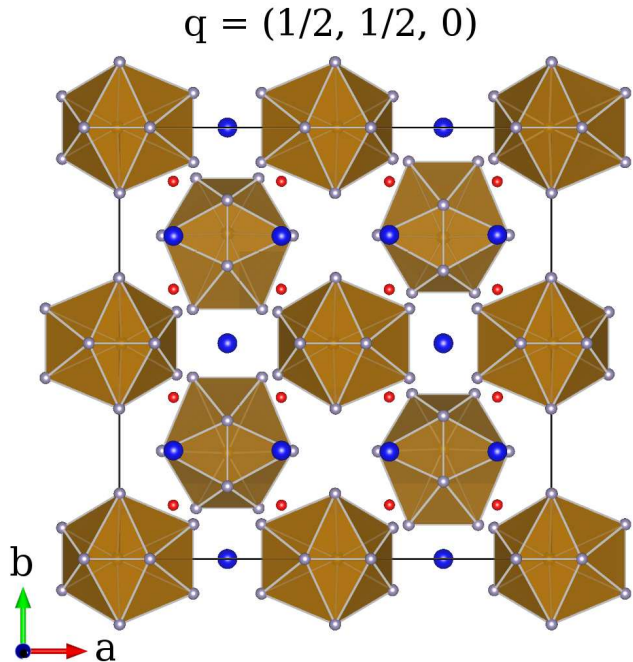}% Here is how to import EPS art
\caption{\label{fig:MMode} (Color online) A 2$\times$2$\times$2 supercell is shown with the lowest energy distortion due to the phonon mode at M. The mode motion that compresses one side of the icosahedra while expanding the other is evident.}
\end{figure}

The distortions due to modes at \textbf{X}=(0.5,0,0) and \textbf{M}=(0.5,0.5,0) give comparable energetic advantages (within the accuracy of DFT) compared to the \textit{I} phase. Single crystal X-ray diffraction indicates a good refinement of the low temperature \textit{I'} structure to a body-centred cubic (BCC) space group $I\bar{4}3d$\cite{Klintberg1}. This favors the mode at (0.5,0.5,0) as the driving structural instability since, along with symmetry equivalent (0.5,0,0.5) and (0,0.5,0.5), the associated distortion produces a face-centred cubic (FCC) cell in reciprocal space. An FCC reciprocal lattice corresponds to a BCC real space lattice, which is in agreement with experiment. In contrast, the distortion due to the low energy mode at \textbf{X} would give rise to simple cubic cells in both reciprocal and real space.

Indeed, when we set into a 2$\times$2$\times$2 supercell a structural modulation that is the superposition of the three symmetry equivalent \textbf{M} modes the symmetry is that of the experimentally observed $I\bar{4}3d$ space group. In this structure there are eight inequivalent atomic sites. The relaxed coordinates in this low temperature distorted structure are shown in Table~\ref{Tab:lowTCoords} for Sr$_3$Ir$_4$Sn$_{13}$. In the high temperature $I$-phase there is only a single free Sn coordinate at the 24$k$ site. In the low temperature structure the phonon distortion results in a free Sr coordinate, two free Ir coordinates and five free Sn coordinates Sn1-Sn5. It is the movement of the Sn coordinates in response to the phonon instabilities that dominates the low energy mode. Fig.~\ref{fig:MMode} depicts the structure when distorted by the mode at M. The primary motion is a breathing of the Sn$_{12}$ cages that compresses one side of each Sn$_{12}$ cage while expanding the other side.

\begin{table}
\small
  \caption{\ Predicted internal ion coordinates for the low temperature structure of Sr$_3$Ir$_4$Sn$_{13}$ modulated by the phonon mode at \textbf{M}. Calculated at the experimental lattice parameters.}
  \label{Tab:lowTCoords}
  \begin{tabular*}{0.45\textwidth}{@{\extracolsep{\fill}}ccc}
    \hline
    Species  & Wyckoff & Position \\
    \hline
    Sr & 48$e$ & (0.2491, 0.4997, 0.1259) \\
    Ir1 & 16$c$ & (0.3747, 0.3747, 0.3747) \\
    Ir2 & 48$e$ & (0.3750, 0.3750, 0.1249) \\
    Sn1 & 16$c$ & (0.0004, 0.0004, 0.0004) \\
    Sn2 & 48$e$ & (0.3267, 0.4012, 0.2500)) \\
    Sn3 & 48$e$ & (0.4235, 0.4997, 0.1506)) \\
    Sn4 & 48$e$ & (0.3494, 0.4237, 0.0002) \\
    Sn5 & 48$e$ & (0.4997, 0.3494, 0.0764) \\
    \hline
  \end{tabular*}
\end{table}

The presence of subtle distortions to the cubic cell, for instance to a tetragonal or monoclinic symmetry as noted in related structures\cite{Gumeniuk1, Gumeniuk2} can not be ruled out. However, it is likely that high intensity synchrotron X-ray diffraction on powdered samples will be required to distinguish superlattice peaks associated with such a distortion, which are beyond the resolution capabilities of the single crystal diffraction experiments performed on this system so far. Nevertheless, in that scenario our conclusion regarding the importance of the phonon mode at \textbf{M} will remain robust.

\subsection{Electronic Structure and Electron-Phonon Coupling}
The preceding results demonstrate that chemical and hydrostatic pressure affect the phonon instabilities, but the lattice degrees of freedom also couple to the electronic structure. The force constants that determine the phononic instabilities are determined via the Hellman-Feynman theorem and therefore depend directly on the electronic charge density and bonding. The clear experimental evidence for superconductivity in (Sr,Ca)$_3$Ir$_4$Sn$_{13}$, highlights the need to understand the evolution of the electronic structure.

\begin{figure}
\includegraphics[scale=0.67,angle=0]{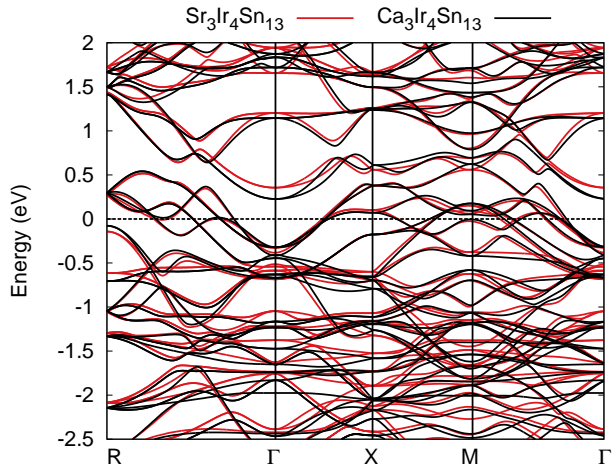}% Here is how to import EPS art
\caption{\label{fig:BandsSr3Ca3} (Color online) Electronic bandstructure of Sr$_3$Ir$_4$Sn$_{13}$ and Ca$_3$Ir$_4$Sn$_{13}$.}
\end{figure}

\begin{figure*}
\includegraphics[scale=0.9,angle=0]{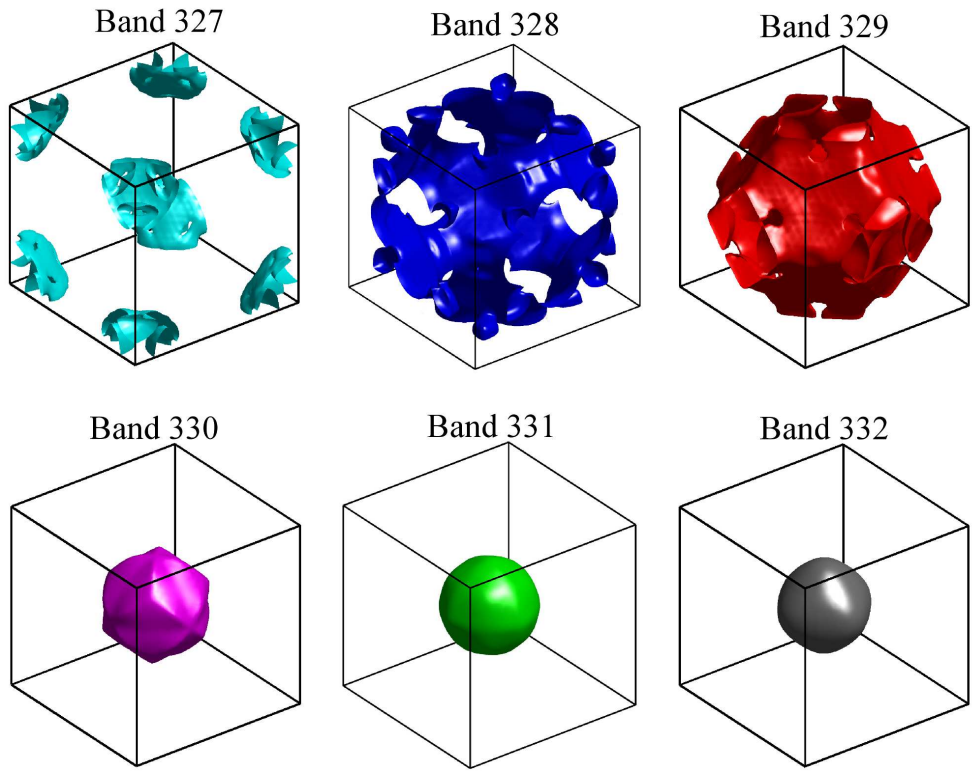}% Here is how to import EPS art
\caption{\label{fig:FS} (Color online) Fermi surfaces of Ca$_3$Ir$_4$Sn$_{13}$.}
\end{figure*}

We have calculated the bandstructure of (Sr,Ca)$_3$Ir$_4$Sn$_{13}$ along high symmetry directions as shown in Fig.~\ref{fig:BandsSr3Ca3}. The bandstructures possess dispersive states near the Fermi level along all directions which is consistent with the cubic structure and metallic conductivity measured experimentally\cite{Klintberg1}. The bandstructures of the two materials are almost perfectly overlaid near the Fermi level and this is reflected in a near identical Fermi surface, which is plotted in Fig.~\ref{fig:FS} for Ca$_3$Ir$_4$Sn$_{13}$. The small surface 327 and the two large sufaces 328 and 328 exhibit flat sections, which may provide opportunities for enhanced superconducting pairing. The 330, 331 and 332 surfaces are more free electron-like and likely to contribute to the good isotropic conductivity of the material in the normal state. 

Away from the Fermi level, the bandstructure for the two materials, (Sr,Ca)$_3$Ir$_4$Sn$_{13}$, differ. For example, inspection of Fig.~\ref{fig:BandsSr3Ca3} at the $\Gamma$ point at -0.55 eV shows a dense clustering of states in Sr$_3$Ir$_4$Sn$_{13}$ that possess a rather flat dispersion. However, the features in this energy range for Ca$_3$Ir$_4$Sn$_{13}$ differ significantly. The clustering of states at -0.55 eV at $\Gamma$ is now split into a less tightly bound set of states near -0.65 eV and a second set closer to -0.45 eV. Furthermore, these states are considerably more dispersive. The bandstructures of the occupied states differ at other points in the Brillouin zone and this will determine the force constants that drive the evolution of the structural instability as the composition changes.

The system's superconductivity is driven by the coupling of the lattice instabilities to the electronic structure. Phonon mediated superconductivity is consistent with recent specific heat\cite{Hayamizu1, Wang1} and thermal conductivity\cite{Zhou1} measurements, which indicate a nodeless gap structure. A full determination of the electron-phonon coupling for all modes in this 40 atom unit cell is beyond the scope of this study and is a challenge to current methods. However, we may obtain a quantitative measure of the electron-phonon (EP) coupling due to a particular mode by evaluating the Fermi-surface averaged
deformation potential\cite{Yildirim1, Khan1} $\Delta = \left\langle \left[ \delta \epsilon(\mathbf{k}) - \delta\mu \right]^2 \right\rangle $. Here, $\delta \epsilon(\mathbf{k})$ is the change in the one-particle energy with momentum $\mathbf{k}$ due to the mode and $\delta\mu$ is the change in chemical potential. $\langle \rangle$ denotes an average of $\mathbf{k}$ over the Fermi surface. We have calculated the deformation potential due to the lowest energy modes at \textbf{X} and \textbf{M} by setting the distortions into a supercell. The unmodulated structure of Ca$_3$Ir$_4$Sn$_{13}$ at $P=35$ kbar, which is beyond complications due to the structural transition, was employed and the mode distortion set in with a Sn displacement $u_{Sn}\approx0.1$ \AA. We obtain a deformation potential of 0.19 eV$^2$ at \textbf{X} and of 0.13 eV$^2$ at \textbf{M} which indicates a strong electron phonon coupling. This suggests that the low energy modes of (Sr,Ca)$_3$Ir$_4$Sn$_{13}$ may be responsible for not only the evolution of the structural transition, but may also be the driver of the competing superconducting phase. However, whether other modes beyond these soft modes are important to the superconductivity requires further theoretical and experimental investigation. The analogy with the numerous examples of emergent superconductivity at magnetic quantum phase transitions is tantalising, and this work invites detailed inelastic X-ray diffraction studies to experimentally identify the key modes in this system of materials. 

\section{Conclusions}
\noindent To conclude the key results include:\\
\noindent (1) Sr$_3$Ir$_4$Sn$_{13}$ possesses imaginary phonons at wavevectors at \textbf{X} ($\mathbf{q}=(0.5,0,0)$), \textbf{M} ($\mathbf{q}=(0.5,0.5,0)$) and \textbf{R} ($\mathbf{q}=(0.5,0.5,0.5)$). \\
\noindent (2) The application of chemical pressure by substitution of Ca onto the Sr site results in the modes at \textbf{X}, \textbf{M} and \textbf{R} evolving towards real frequencies. Calculation of total energies for supercells incorporating the three lowest lying modes at \textbf{X}, \textbf{M} and \textbf{R} reveals that the modes at \textbf{X} and \textbf{M} give energy advantages of 4.6 meV/F.U. and 2.1 meV/F.U. compared to the high temperature \textit{I} phase. This is consistent with the transition temperature of 33 K for the structural transition in the Ca substituted compound. The space group symmetry from the experimental X-ray refinement to a body-centred cubic structure below the transition temperature corresponds to the instability at \textbf{M} ($\mathbf{q}=(0.5,0.5,0)$). \\
\noindent (3) Applying a pressure of 35 kbar to Ca$_3$Ir$_4$Sn$_{13}$ in our simulations results in the removal of all instabilities towards these phonon modes. This is consistent with the end point of the structural phase transition at approximately 18 kbar in the experimental temperature-pressure phase diagram. This implies a role for the modes at \textbf{X} and \textbf{M} in the structural transition and superconductivity of this system. \\ 
\noindent (4) The presence of low energy phonon modes that may be tuned to zero energy by the application of pressure supports the identification of a quantum phase transition in this system, which may be second order.\\
\noindent (5) Finally, the calculated deformation potentials due to low energy modes in this system suggests a phonon mediated mechanism for its superconductivity.\\

%\noindent (5) The Fermi surfaces of Sr$_3$Ir$_4$Sn$_{13}$ and Ca$_3$Ir$_4$Sn$_{13}$ possess a 3D morphology with flat sections. Calculation of the real part of the susceptibility for band 328 indicates enhancement of 21\% at \textbf{X} over its value at the Brillouin zone center. This enhancement may couple to the demonstrated phonon instabilities to play a role in the structural transition and superconductivity.

\begin{acknowledgments}
I acknowledge discussions with S.K. Goh, F.M. Grosche and P.J. Saines and financial support of the EPSRC, UK.
\end{acknowledgments}

% The \nocite command causes all entries in a bibliography to be printed out
% whether or not they are actually referenced in the text. This is appropriate
% for the sample file to show the different styles of references, but authors
% most likely will not want to use it.
%\nocite{*}

\bibliography{Sn13}% Produces the bibliography via BibTeX.

\end{document}